\documentclass[letterpaper, 11 pt, paper]{article}  

\usepackage{graphicx}                                                          
\usepackage{amsmath,amssymb}                                                         

\title{\LARGE \bf
Nonlinear Robust Filtering of Sampled-Data Dynamical Systems}
\date{}

\author{Masoud Abbaszadeh$^{*}$ and Horacio J. Marquez
\thanks{M. Abbaszadeh is with GE Global Research, Niskayuna, NY, USA 12309,
        {\tt\small The corresponding author, e-mail: masoud@ualberta.net}}%
\thanks{H. J. Marquez is with the Department of Electrical and Computer Engineering,
        University of Alberta, Edmonton, Alberta, Canada, T6G 2V4,
        {\tt\small e-mail: marquez@ece.ualberta.ca}}%
}

\begin{document}
\maketitle

\begin{abstract}
This work is concerned with robust filtering of nonlinear sampled-data systems with and
without exact discrete-time models. A linear matrix inequality (LMI)
based approach is proposed for the design of robust $H_{\infty}$
observers for a class of Lipschitz nonlinear systems. Two type of
systems are considered, Lipschitz nonlinear discrete-time systems
and Lipschitz nonlinear sampled-data systems with Euler approximate
discrete-time models. Observer convergence when the exact
discrete-time model of the system is available is shown. Then,
practical convergence of the proposed observer is proved using the
Euler approximate discrete-time model. As an additional feature,
maximizing the admissible Lipschitz constant, the solution of the
proposed LMI optimization problem guaranties robustness against some
nonlinear uncertainty. The robust $H_{\infty}$ observer synthesis
problem is solved for both cases. The maximum disturbance
attenuation level is achieved through LMI optimization.
At the end, a path to extending the results to higher-order
approximate discretizations is provided.
\end{abstract}


\section{Introduction}
Design of discrete-time nonlinear observers has been
the subject of significant attention in recent years.
\cite{Califano}, \cite{Xiao1}, \cite{Kazantzis}, \cite{Wang}. The
study of the nonlinear discrete-time observers is important at least
for two reasons. First, most continuous-time control system designs
are implemented digitally. Given that in most practical cases it is
impossible to measure every state variable in real time, these
controllers require the reconstruction of the states of the
discrete-time model of the true continuous-time plant. Second, there
are systems which are inherently discrete-time and do not originate
from discretization of a continuous-time plant. Of those,
discrete-time observers of continuous-time systems are particularly
challenging. The reason is that exact discretization of a
continuous-time nonlinear model is usually not possible to obtain.
Approximate discrete-time models, on the other hand, are affected by
the consequent approximation error. In this paper, we address both
problems. First, we consider a class of nonlinear discrete-time
systems with exact model. A nonlinear $H_{\infty}$ observer design
algorithm is proposed for these systems based on an LMI approach.
Then, the nonlinear sampled data system with Euler approximate model
is considered. The Euler approximation is important because not only
it is easy to derive but also maintains the structure of the
original nonlinear model. We will show that by appropriate selection
of one of the parameters in our proposed LMIs (actually the only
design parameter in our algorithm), the practical convergence of the
observer via Euler approximation is guaranteed as well as the robust
$H_{\infty}$ cost. Our approach is based on the recent results of
\cite{Arcak}. See \cite{Hammouri} and \cite{Saif} for other
approaches. We emphasize that while the algorithms in
\cite{Hammouri} and \cite{Saif} are specifically designed for Euler
discretization, our proposed algorithm can be applied either to the
nominal exact discrete-time model or its Euler approximation.

There is a large body of literature for control and estimation of nonlinear systems satisfying a Lipschitz continuity condition.
See for example \cite{abbaszadeh2008robust, abbaszadeh2007robust, Raghavan, abbaszadeh2006robust,Xu1, Xu2, Xu3, Xu4,Hammouri,Lu,Gao,
abbaszadeh2010nonlinear, abbaszadeh2008lmi, Thau1, abbaszadeh2010dynamical, abbaszadeh2010robust2,Rajamani2, deSouza1, deSouza2, abbaszadeh_phdthesis,
abbaszadeh2012generalized,Abbaszadeh5} and the references therein, for details of the approach and applications to
control and filtering of different classes of nonlinear systems. The significance of this condition is that it guarantees the existence and uniqueness of the solution of the nonlinear systems. Also, it provides a mathematically tractable framework to apply Lyapunov stability theory and establish stability and performance conditions in the form of Riccati equations or LMIs.

The LMI based observer design for uncertain discrete-time systems has
been addressed in several works e.g. \cite{Xu1}, \cite{Lu} and
\cite{Xu2}. In all these studies, the proposed LMIs are nonlinear in
the Lipschitz constant and thus it can not be considered as one of
the LMI variables. In the algorithm proposed here, first the problem
is addressed in the general case, then, having a bound on the
Lipschitz constant, the LMIs become linear in the Lipschitz constant
and we can take advantage of this feature to solve an optimization
problem over it. Providing that the optimal solution is larger than
the actual Lipschitz constant of the system in hand, we show that
the redundancy achieved can guarantee robustness against some
nonlinear uncertainty in the original continuous-time model for both
exact and Euler approximate discretizations.

The rest of the paper is organized
as follows: Section II briefly describes the filtering framework.
In Section III, an observer design method for a class of
nonlinear discrete-time systems is introduced. In Section IV the
practical convergence of the proposed observer via the Euler
approximate models is shown. In section V, the results of the two
previous sections will extend into the $H_{\infty}$ context followed
by an illustrative example showing satisfactory performance of our
algorithm.

\section{Filtering Framework}
\begin{figure}[!h]
  \centering
  \includegraphics[trim= 3mm 80mm 3mm 15mm, clip, width=\textwidth]{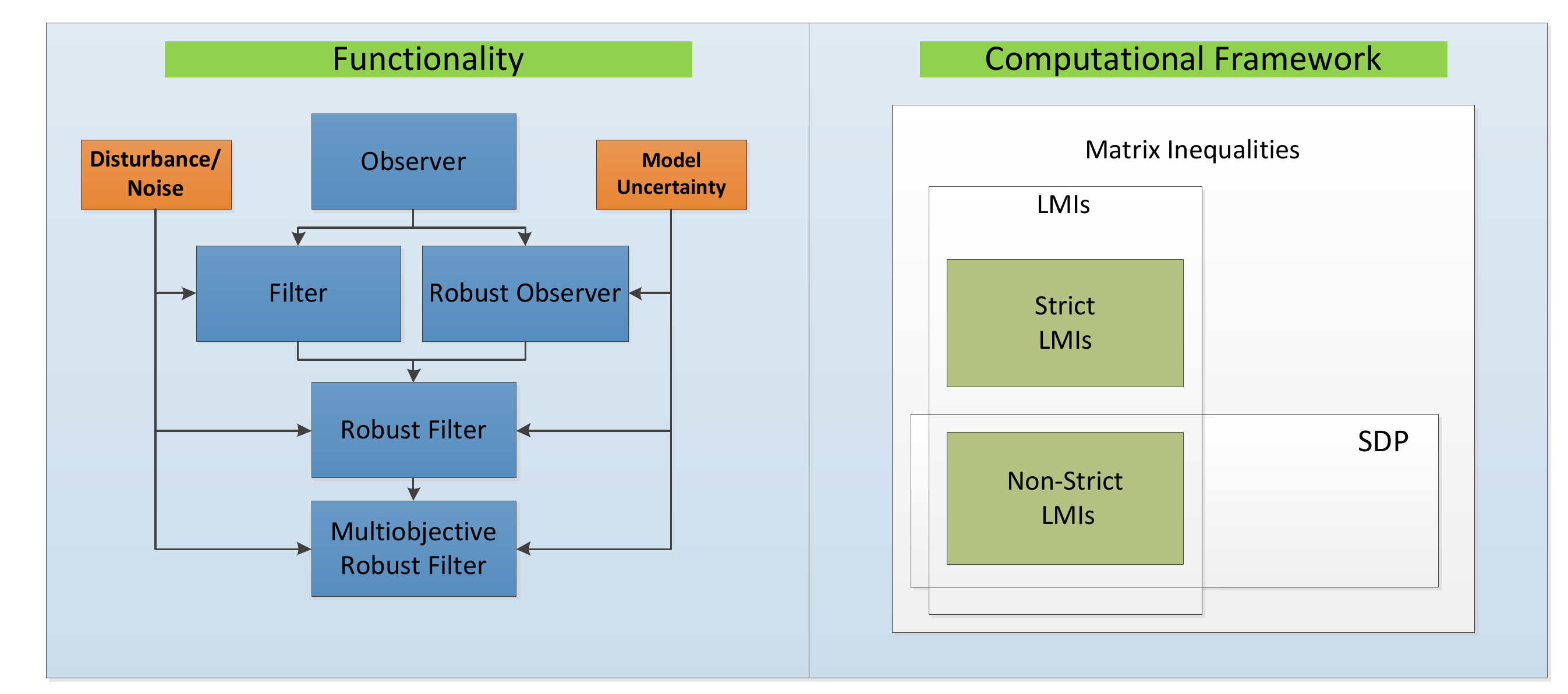}\\
  \caption{State estimation functionality and computational framework}\label{classify_fig}
\end{figure}

Figure \ref{classify_fig} shows a classification of state estimators in terms of their functionality,
and their computational framework \cite{abbaszadeh2014generalized}. The \emph{filtering} problem deals with state estimation under noise/disturbance; the \emph{robust observation} problem addresses state estimation under model uncertainty, while \emph{robust filtering} combines the two. \emph{Multi-objective robust filtering} provides tools to tune the trade-offs between robustness bounds, disturbance attenuation level and convergence rate \cite{abbaszadeh_phdthesis, abbaszadeh2010robust2,abbaszadeh2007robust}. While general matrix inequalities, including bilinear matrix inequalities (BMIs), are not numerically tractable, semidefinite programming problems (SDP) and LMIs can be solved using efficient interior-point methods. \emph{Strict LMI}s are referred to those LMIs in which all inequalities are strictly positive or negative definite and no semidefinite matrices or equality constraints are allowed. Strict LMI solvers are often more efficient than SDP solvers. The solutions provided in this work are \emph{robust filters} whose gains are computed using \emph{strict LMI}s. Following an $\mathcal{L}_{2}$ filtering framework, we assume that the noise is energy-bounded and the model uncertainties are norm-bounded.

\section{Observer Design For Nonlinear Discrete-Time Systems}

We consider the following system
\begin{align}
x(k+1)&=A_{d}x(k)+F(x(k),u(k))\label{sys1-1}\\
y(k)&=C_{d}x(k)\label{sys1-2}
\end{align}
where  $x\in {\mathbb R} ^{n} ,u\in {\mathbb R} ^{m} ,y\in {\mathbb
R} ^{p} $ and $F(x(k),u(k))$ contains nonlinearities of second order
or higher. The above system can be either an inherently
discrete-time system or the exact discretization of a
continuous-time system. We assume that $F(x(k),u(k))$  is locally
Lipschitz with respect to $x$ in a region $\mathcal{D}$, uniformly
in $u$, i.e. $\forall \, x_{1} (k),x_{2} (k)\in \mathcal{D}$:
\begin{eqnarray}
\|F(x_{1},u^{*})-F(x_{2},u^{*})\| \le \gamma _{d} \| x_{1}-x_{2}\|
\end{eqnarray}
where $\|.\|$ is the induced 2-norm, $u^{*}$ is any admissible
control sequence and $\gamma_{d}>0$ is called the Lipschitz
constant. If the nonlinear function $F$ satisfies the Lipschitz
continuity condition globally in $\mathbb{R}^{n}$, then all the
results in this and the ensuing sections will be valid globally.
All matrices and vectors have appropriate dimensions unless
otherwise mentioned. The proposed observer is in the following
form:
\begin{equation}
\hat{x} (k+1)=A_{d} \hat{x} (k)+F(\hat{x} (k),u(k))+L(y(k)-C_{d}
\hat{x} (k))\label{observer1}
\end{equation}%
the observer error is thus:
\begin{eqnarray}
\begin{split}
e(k+1) \triangleq& \ x(k+1)-\hat{x}(k+1)=(A_{d}
-LC_{d})e(k)\\&+F(x(k),u(k))-F(\hat{x} (k),u(k)).\label{error1}
\end{split}
\end{eqnarray}
Our goal is two-fold: (i) In the first place, we want to find an
observer gain, $L$, such that the observer error dynamics is
asymptotically stable. (ii) We want to maximize $\gamma _{d}$, the allowable Lipschitz constant of the nonlinear system.\\

\emph{\textbf{Theorem 1.} Consider the system
(\ref{sys1-1})-(\ref{sys1-2}) with given Lipschitz constant
$\gamma _{d}$. The observer error dynamics (\ref{error1}) is
(globally) asymptotically stable if there exist scalar
$\varepsilon>0$, fixed
matrix $Q>0$ and matrices $P>0$ and $G$ such that the following set of LMIs has a solution\\
\begin{align}
&\left[
\begin{array}{cc}
P-Q-\varepsilon I & A_{d}^{ T} P-C_{d}^{T} G^{T} \\
PA_{d} -GC_{d} & P%
\end{array}
\right] >0 \\
&\left[
\begin{array}{cc}
\Psi_{1}I & P \\
P & \Psi_{1}I %
\end{array}
\right] >0
\end{align}
where
\begin{equation}
\Psi_{1}=\frac{-\lambda _{\max } (Q)+\sqrt{\lambda _{\max }^{2}
(Q)+\frac{1}{\gamma _{d}^{2}}\lambda_{\min}^{2}(Q)} }{\gamma
_{d}+2}.\label{si1}
\end{equation}
$P$, $G$, and $\epsilon$ are the LMI variables and $Q$ is a design
parameter to be chosen. Once the problem is solved:}
\begin{gather}
L=P^{-1}G
\end{gather}

\textbf{Proof: }Consider the Lyapunov function candidate as
follows:
\begin{equation}
V_{k} =e_{k}^{T} Pe_{k}
\end{equation}
then:
\begin{eqnarray}
\begin{split}
{\Delta V} &=V_{k+1} -V_{k} =e_{k}^{T} (A_{d} -LC_{d} )^{T} P(A_{d} -LC_{d})e_{k}\\
&\ \ \ +2e_{k}^{T}(A_{d} -LC_{d} )^{T} P(F_{k} -\hat{F}_{k} )\\
& \ \ \
+(F_{k}-\hat{F}_{k})^{T}P(F_{k}-\hat{F}_{k})-e_{k}^{T}Pe_{k}\label{Lyap1}
\end{split}
\end{eqnarray}
where for simplicity:
\begin{eqnarray}
F_{k} \triangleq F(x(k),u(k)),\, \hat{F}_{k} \triangleq
F(\hat{x}(k),u(k)).
\end{eqnarray}
Suppose  $\exists \, P,Q>0$  such that the following discrete-time
Lyapunov equation has a solution:
\begin{equation}
\displaystyle (A_{d} -LC_{d} )^{T} P(A_{d} -LC_{d}
)-P=-Q\label{Lyap2}
\end{equation}
then (\ref{Lyap1}) becomes:
\begin{equation}
\begin{split}
\Delta V=&-e_{k}^{T} Qe_{k} +2e_{k}^{T} (A_{d} -LC_{d} )^{T} P(F_{k}
-\hat{F}_{k} )\\
&+(F_{k} -\hat{F}_{k} )^{T} P(F_{k} -\hat{F}_{k})
\end{split}
\end{equation}
using Rayleigh and Schwartz inequalities, we have:
\begin{align}
&\|e_{k}^{T} Qe_{k}\| \ge \lambda _{\min }(Q)\|e_{k}\| ^{2}\\
&\|2e_{k}^{T} (A_{d} -LC_{d} )^{T} P(F_{k} -\hat{F}_{k} )\| \le \|
2e_{k}^{T} P(F_{k} -\hat{F}_{k} )\|\cdots \notag \\
& \ \ \ \cdots \| A_{d} -LC_{d} \| \le 2\gamma_{d} \lambda _{\max }
(P)\|
e_{k} \| ^{2} \| A_{d} -LC_{d}\|\notag \\
& \ \ \ =2\gamma _{d} \lambda _{\max }(P)\|e_{k} \| ^{2} \bar{\sigma}(A_{d} -LC_{d})\label{f1}\\
&\| (F_{k} -\hat{F}_{k} )^{T} P(F_{k} -\hat{F}_{k} )\| \le \lambda
_{\max } (P)\| (F_{k} -\hat{F}_{k} )\| ^{2}\notag \\
& \ \ \ \le \gamma _{d}^{2} \lambda _{\max } (P)\| e_{k}
\|^{2}\label{f2}
\end{align}
so for  $\Delta V<0$  it is sufficient to have:
\begin{equation}
-\lambda _{\min } (Q)+\lambda _{\max } (P)[2\gamma _{d}\bar{\sigma
}(A_{d} -LC_{d} )+\gamma _{d}^{2}]<0 \label{cond1}.
\end{equation}
Condition (\ref{cond1}) along with (\ref{Lyap2}) are sufficient
conditions for asymptotic stability. We now endeavor to convert
these nonlinear inequalities into LMIs. There exists a solution
for (\ref{Lyap2}) if
\begin{multline}
\exists \ \varepsilon >0,\, \, (A_{d} -LC_{d} )^{T} P(A_{d}-LC_{d}
)-P<-Q-\varepsilon I \\
\Rightarrow(P-Q-\varepsilon I)-(A_{d} -LC_{d} )^{T} PP^{-1}
P(A_{d} -LC_{d})>0
\end{multline}
using Schur's complement lemma, defining  $G=PL$ and knowing that
$P^{T}=P$, the first LMI in Theorem 1 is obtained.
 The Lyapunov equation in
(\ref{Lyap2}) can be rewritten as
\begin{equation}
P=(A_{d} -LC_{d} )^{T} P(A_{d} -LC_{d} )+Q
\end{equation}
and taking into account that:
\begin{equation}
\left|\bar{\sigma }(A)-\bar{\sigma }(B)\right|\le \bar{\sigma
}(A+B)\le \bar{\sigma }(A)+\bar{\sigma }(B)
\end{equation}
we have that
\begin{eqnarray}
\begin{split}
&|\bar{\sigma}\left[(A_{d} -LC_{d} )^{T} P(A_{d} -LC_{d}
)\right]-\bar{\sigma }(Q)|\le \bar{\sigma}(P)\\
&\Rightarrow \bar{\sigma }\left[(A_{d} -LC_{d} )^{T} P(A_{d} -LC_{d}
)\right]\le \bar{\sigma }(Q)+\bar{\sigma }(P)\label{ineq1}
\end{split}
\end{eqnarray}
using Schwartz inequality:
\begin{equation}
\displaystyle \bar{\sigma }\left[(A_{d} -LC_{d} )^{T} P(A_{d}
-LC_{d} )\right]\le \bar{\sigma }^{2} (A_{d} -LC_{d}
)\bar{\sigma}(P)\label{ineq2}
\end{equation}
comparing (\ref{ineq1}) and (\ref{ineq2}), a sufficient condition
for (\ref{ineq1}) is
\begin{multline}
\bar{\sigma }^{2} (A_{d} -LC_{d})\bar{\sigma }(P)\le \bar{\sigma}(P)+\bar{\sigma }(Q)\\
\Rightarrow \bar{\sigma }(A_{d} -LC_{d} )\le
\sqrt{1+\frac{\bar{\sigma }(Q)}{\bar{\sigma }(P)} }\label{ineq3}
\end{multline}
note that since {\it P} and {\it Q} are positive definite their
eigenvalues and singular values are the same. Now, we want to find
a sufficient condition for (\ref{cond1}). Using (\ref{ineq3}):
\begin{equation}
\begin{split}
\bar{\sigma }(A_{d} -LC_{d} )\lambda_{\max } (P)&\le
\sqrt{1+\frac{\bar{\sigma }(Q)}{\bar{\sigma}(P)} } \bar{\sigma }(P) \\
&=\sqrt{\bar{\sigma }^{2} (P)+\bar{\sigma }(Q)\bar{\sigma }(P)}.
\end{split}
\end{equation}
Suppose {\it Q} is given, define,
\begin{equation}
\displaystyle \begin{array}{l} {g(\bar{\sigma }(P))\triangleq
\sqrt{\bar{\sigma }^{2} (P)+\bar{\sigma }(Q)\bar{\sigma }(P)}}
\end{array}
\end{equation}
then
$g(\bar{\sigma }(P))$  is strictly increasing so there is no
constant upper limit for this function but we can still bound this
nonlinear function with a linear one.
\begin{align}
&g(\bar{\sigma }(P)) <\sqrt{\bar{\sigma }^{2} (P)+\bar{\sigma}(Q)\bar{\sigma }(P)+\left[\frac{\bar{\sigma }(Q)}{2} \right]^{2}} \notag \\
&\hspace{1.2cm}=\sigma(P)+\frac{\bar{\sigma }(Q)}{2}\label{sigma1}\\
&\Rightarrow\bar{\sigma}(A_{d}-LC_{d})\lambda_{\max}(P)<\bar{\sigma}(P)+\frac{\bar{\sigma
}(Q)}{2}\label{ineq6}
\end{align}
which is a sufficient condition for (\ref{ineq1}). Substituting
the above into (\ref{cond1}), a sufficient condition for
(\ref{cond1}) is
\begin{equation}
\gamma _{d}\lambda _{\max } (P)\left[2\sqrt{1+\frac{\lambda _{\max
} (Q)}{\lambda _{\max } (P)}}+\gamma_{d}\right] <\lambda _{\min }
(Q).\\\label{lambda1}
\end{equation}
For any $a,b>0$, $a^{2}<b^{2}$ implies $a<b$, thus, by squaring the
two sides of the above inequality, substituting from (\ref{sigma1})
and after some algebra, to have (\ref{lambda1}) it suffices to
\begin{multline}
(\gamma_{d}+2)\lambda_{\max}^{2}(P)+2\lambda_{\max}(Q)\lambda_{\max}(P)<\frac{\lambda_{\min}^{2}(Q)}{\gamma_{d}^{2}(\gamma_{d}+2)}\\
\Rightarrow \lambda_{\max}(P)<\Psi_{1}\label{lambda2}
\end{multline}
or equivalently,
\begin{equation}
\Psi_{1}^{2}I-PP^{T}>0
\end{equation}
which is by means of Schur's complement lemma, equivalent to the
second LMI in Theorem 1 where $\Psi_{1}$ is as in \eqref{si1}. This
ends
the proof. $\triangle$\\

In continuance, consider the case where the Lipschitz constant of
system, $\gamma _{d}$, is less than 1. This is not restrictive since
the Lipschitz constant can reduced using a suitable coordinate
transformation \cite{Raghavan}. Besides, the discretized models of
continuous-time systems may also fall into this category by
appropriate selection of the sampling time. We will see this in
detail for Euler discretization in the next section. The following
theorem shows that under this assumption, the maximum admissible
Lipschitz constant is achievable through an LMI optimization over $\gamma _{d}$.\\

\emph{\textbf{Theorem 2.} Consider the system
(\ref{sys1-1})-(\ref{sys1-2}). The observer error dynamics
(\ref{error1}) is (globally) asymptotically stable with maximum
admissible Lipschitz constant $\gamma _{d}^{\ast}$, if there exist
scalars $\varepsilon>0,\xi>1$, fixed matrix $Q>0$ and matrices $P>0$
and $G$ such that the following LMI optimization problem has a
solution
\begin{eqnarray}
\min (\xi )\notag
\end{eqnarray}
\hspace{1.5cm}s.t.
\begin{align}
&\left[
\begin{array}{cc}
P-Q-\varepsilon I & A_{d}^{ T} P-C_{d}^{T} G^{T} \\
PA_{d} -GC_{d} & P%
\end{array}\label{LMI2}
\right] >0 \\
&\left[
\begin{array}{cc}
\Psi_{2}I & P \\
P & \Psi_{2}I%
\end{array}
\right] >0 \label{LMI1}
\end{align}\\
where
\begin{equation}
\Psi_{2}=\dfrac{1}{3} \left[ \lambda _{\min } \left( Q\right) \xi
-\lambda _{\max } \left( Q\right) \right].
\end{equation}
once the problem is solved:}
\begin{align}
L&=P^{-1}G \\
\gamma _{d}^{\ast}&\triangleq\max (\gamma _{d} )=\dfrac{1}{\xi }
\end{align}

\textbf{Proof: } Having the same Lyapunov function candidate it
follows that, $\triangle V$ is given by (\ref{Lyap1}). Knowing
$\gamma_{d}<1$, \eqref{cond1} reduces to
\begin{equation} \gamma
_{d} <\frac{\lambda _{\min } (Q)}{\left[2\bar{\sigma }(A_{d}
-LC_{d} )+1\right]\lambda _{\max } (P)}\label{gamma1}
\end{equation}
where $Q$ is the same as before. Based on (\ref{ineq6}), it can be
written
\begin{equation}
\left[2\bar{\sigma }(A_{d} -LC_{d} )+1\right]\lambda_{\max }
(P)<3\bar{\sigma }(P)+\bar{\sigma }(Q).
\end{equation}
From the above, we have
\begin{align}
&\frac{\lambda _{\min } (Q)}{\left[2\bar{\sigma }(A_{d} -LC_{d}
)+1\right]\lambda _{\max } (P)} >\frac{\lambda _{\min }
(Q)}{3\bar{\sigma }(P)+\lambda _{\max } (Q)}\notag.
\end{align}
Eventually, a sufficiency condition for (\ref{gamma1}) is
\begin{equation}
\gamma _{d} <\frac{\lambda _{\min } (Q)}{3\bar{\sigma }(P)+\lambda
_{\max } (Q)} \to \bar{\sigma }(P)<\frac{\lambda _{\min }
(Q)}{3\gamma _{d} } -\frac{1}{3} \lambda _{\max }
(Q)\label{lambda3}
\end{equation}
which, by means of Schur's complement lemma is equivalent to the second LMI in Theorem 2. $\triangle$\\

\emph{\textbf{Remark 1.}} The purpose of Theorem 2 is two-fold. (i)
to find a gain matrix ``$L$" that stabilizes the observer error
dynamics, and (ii) to maximized $\gamma_{d}$. Dropping the
maximization of $\gamma_{d}$ still renders a stable observer. In
this case the proposed LMI optimization reduces to an LMI
feasibility problem (namely; satisfying the constraints) which is
easier. The only parameter to be chosen in both cases is the positive definite matrix {\it Q}.\\

\emph{\textbf{Remark 2- Nonlinear Uncertainty. }}The advantage of
maximization of $\gamma_{d}$ is that if the maximum admissible
Lipschitz constant achieved by Theorem 1, $\gamma _{d}^{*} $, is
greater than the actual Lipschitz constant of the system, $\gamma
_{d} $, then the proposed observer can tolerate some nonlinear
uncertainty. Consider the system with nonlinear uncertainty as
below:
\begin{align}
F_{\Delta } (x,u)&\triangleq F(x,u)+\Delta F(x,u)\\
x(k+1)&= A_{d} x(k)+F_{\Delta } (x,u) \\
y(k)&= C_{d} x(k).
\end{align}
Suppose the additive nonlinear uncertainty is Lipschitz with unknown
Lipschitz constant  $\Delta \gamma _{d}$. According to the Theorem
1, $F_{\Delta } (x(k),u(k))$ can be any Lipschitz nonlinear function
with Lipschitz constant less than or equal to $\gamma _{d}^{*} $,
i.e.:
\begin{equation}
\| F_{\Delta } (x_{1} ,u)-F_{\Delta } (x_{2} ,u)\| \le
\gamma_{d}^{*} \| x_{1} -x_{2} \|.
\end{equation}
On the other hand
\begin{equation}
\begin{split}
&\| F_{\Delta } (x_{1} ,u)-F_{\Delta } (x_{2} ,u)\|=\|F(x_{1},u)+\Delta F(x_{1} ,u)\cdots\\&\cdots-F(x_{2} ,u)-\Delta F(x_{2} ,u)\| \notag\\
&\le \| F(x_{1} ,u)-F(x_{2} ,u)\| +\| \Delta F(x_{1}
,u)-\Delta F(x_{2} ,u)\| \notag\\
&\le \gamma _{d} \| x_{1} -x_{2}\| +\Delta \gamma _{d} \| x_{1}
-x_{2} \|
\end{split}
\end{equation}
So, there must be:
\begin{equation}
\displaystyle \gamma _{d} +\Delta \gamma _{d} \le \gamma _{d}^{*}
\to \Delta \gamma _{d} \le \gamma _{d}^{*} -\gamma _{d}.
\end{equation}
This means that the proposed observer is robust against any
additive Lipschitz nonlinear uncertainty with Lipschitz constant
less than or equal to $\gamma _{d}^{*} -\gamma _{d}$.

\section{Observer Design For Nonlinear Sampled-Data Systems Via Euler Approximation}

In usual, given a continuous nonlinear model, an exact
discretization can not be found in closed form, thus originating the
need of approximate discrete-time models. A framework for nonlinear
observer design based on approximated models has been recently
proposed in \cite{Arcak}. In this section, our focus will be on
Euler approximation which is an important case because it is easy to
derive and it doesn't change the structure of the original nonlinear
model. Following the notation of \cite{Arcak}, we consider the
following continuous-time system
\begin{eqnarray}
\displaystyle \begin{array}{l} {\dot{x}=Ax+f(x,u)} \\ {y=Cx}
\end{array}\label{dis1}
\end{eqnarray}
where $x\in {\mathbb R} ^{n} ,u\in {\mathbb R} ^{m} ,y\in {\mathbb
R} ^{p} $. We assume that system has an equilibrium point at the
origin and $f(x,u)$ is locally Lipschitz with the Lipschitz
constant $\gamma _{c} $. The family of exact discretizations of
(\ref{dis1}) is:
\begin{eqnarray}
\displaystyle \begin{array}{l} {x(k+1)=A_{d} x(k)+F_{T}^{e} (x(k),u(k))} \\
{y(k)=C_{d} x(k)} \end{array}\label{dis2}
\end{eqnarray}
index {\it T} means the discretization is dependent to the sampling
time, {\it T}. To compute (\ref{dis2}) we need a closed-form
solution of (\ref{dis1}) over the sampling intervals {\it [k
T,(k+1)T)}, which is hard to obtain or even impossible. However, it
is realistic to assume that a family of approximate discrete-time
models is available
\begin{eqnarray}
\displaystyle \begin{array}{l} {x^{a} (k+1)=A_{d}^{a}
x(k)+F_{T}^{a} (x(k),u(k))} \\ {y(k)=C_{d} x^{a} (k)}.
\end{array}\label{dis3}
\end{eqnarray}
Then for the Euler approximation we have
\begin{eqnarray}
A_{d}^{a} &=&I+AT \\
F_{T}^{a} (x^{a} (k),u(k))&=& T f(x^{a} (k),u(k))\label{euler1}.
\end{eqnarray}
Similar to (\ref{observer1}), the proposed observer is
\begin{eqnarray}
\displaystyle \hat{x}^{a} _{k+1}=A_{d}^{a} \hat{x}^{a}_{k}+F_{T}^{a}
(\hat{x}^{a}_{k},u_{k})+L(y_{k}-C_{d}
\hat{x}^{a}_{k})\label{observer2}.
\end{eqnarray}
Before expressing our result, we recall two aspects from
\cite{Arcak}, \emph{consistency} and \emph{semiglobal practicality}.
The definitions are omitted here due to space limitations. According
to the verifiable consistency conditions given in \cite{Kokotovic},
if the trajectories of a continuous-time Lipschitz nonlinear system
are bounded, then the Euler approximation is consistent (one-step
consistent) with the exact discrete-time model.\\



Based on (\ref{euler1}), the Lipschitz constant of the Euler
approximation is $\gamma _{d} =T\gamma _{c}$. Again, we assume
$\gamma _{d} <1$. This is even less restrictive than in section 2,
because here {\it T} directly multiplies $\gamma _{c} $ and can be
chosen sufficiently small. The following theorem shows that how the
algorithm proposed in Theorem 2 can be used to design an observer
using Euler approximate discrete-time model guaranteing observer
practical convergence when applied to the (unknown) exact model, by
the appropriate selection of Q.\\

\emph{\textbf{Theorem 3.} The observer (\ref{observer2}) designed
using the Euler approximate model (\ref{euler1}) is semiglobal
practical in T with the maximum admissible Lipschitz constant
$\gamma _{d}^{\ast}$, if the trajectories of (\ref{dis1}) are
bounded and there exist scalars $\varepsilon >0,\, \xi >1$, fixed
matrix $Q>0$ and matrices $P>0$ and $G$ such that the LMI
optimization problem (\ref{LMI2})-(\ref{LMI1}) has
a solution where $\lambda _{\min } \left(Q\right)=T$.}\\

\textbf{Proof.} Consider the same Lyapunov function used in Theorems
1 and 2:
\begin{equation}
V_{T} (e(k))= \left[x^{a} (k)-\hat{x}^{a} (k)\right]^{a}
P\left[x^{a} (k)-\hat{x}^{a} (k)\right]
\end{equation}
then
\begin{equation}
\begin{split}
&\left\| V_{T} (e_{k_{1}})-V_{T} (e_{k_{2}}))\right\|= \left\| e^{T}
_{k_{1}}Pe_{k_{1}}-e^{T}_{k_{2}}Pe_{k_{2}}\right\|\\
&={\left\| \left[e(k_{1} )-e(k_{2} )\right]^{T} P\left[e(k_{1})+e(k_{2} )\right]\right\| } \\
&\leq \lambda _{\max } (P)\| e(k_{1} )+e(k_{2}
)\|\|e(k_{1})-e(k_{2})\|
\end{split}
\end{equation}
by the definition of observer error, the observer error is finite
(note that the convergence of the observer states to the states of
the Euler approximate model has already been achieved by virtue of
Theorem 2, so
\begin{equation}
\displaystyle \begin{array}{l} {\exists \, M\in \left(0,\infty
\right),\, \, \, \lambda _{\max } (P)\left\| e(k_{1} )+e(k_{2}
)\right\| \le M} \end{array}\end{equation} thus, from the above:
\begin{equation}
\left\| V_{T} (e(k_{1} ))-V_{T} (e(k_{2} ))\right\| \le M\left\|
e(k_{1} )-e(k_{2} )\right\|
\end{equation}
Similar to what we did it section 2, we have
\begin{equation}
F_{k,T}^{a} \triangleq F_{T}^{a} (x^{a} (k),u(k)),\,
\hat{F}_{k,T}^{a} \triangleq F_{T}^{a} (\hat{x}^{a}(k),u(k))
\end{equation}
\begin{equation}
\begin{split}
&V_{T} (e_{k+1})-V_{T} (e_{k})=-e_{k}^{T} Qe_{k}+2e_{k}^{T}(A_{d}-LC_{d} )^{T}\cdots\\
&\cdots P(F_{k,T}^{a}-\hat{F}_{k,T}^{a} )+(F_{k,T}^{a}
-\hat{F}_{k,T}^{a} )^{T} P(F_{k,T}^{a}-\hat{F}_{k,T}^{a}) \\
&\leq -\lambda _{\min }(Q)\| e_{k} \| ^{2}+2\bar{\sigma}(A_{d}^{a}
-LC_{d} )\lambda _{\max } (P)\gamma _{d} \| e_{k} \| ^{2}\\
&\hspace{3.7mm}+\gamma _{d}^{2} \lambda _{\max } (P)\| e_{k} \|^{2}.
\end{split}\label{ineq4}
\end{equation}
Using (\ref{gamma1}) and (\ref{ineq3}), it can be written
\begin{multline}
[-\lambda _{\min } (Q)+2\bar{\sigma }(A_{d}^{a}
-LC_{d} )\lambda _{\max } (P)\gamma _{d}]\| e_{k} \| ^{2}\\
\leq -\frac{\lambda _{\min } (Q)\| e_{k}\| ^{2}}{2\bar{\sigma
}(A_{d} -LC_{d})+1}\le -\frac{\lambda _{\min } (Q)\| e_{k} \|
^{2}}{2\sqrt{1+\frac{\lambda _{\max } (Q)}{\lambda _{\max } (P)}}+1}
\label{ineq5}
\end{multline}
substituting (\ref{ineq5}) into (\ref{ineq4}) and knowing that
$\lambda _{\min } (Q)=T$ and $\gamma_{d}=T\gamma_{c}$, we will have
\begin{equation}
\begin{split}
&{\frac{V_{T} (e(k+1))-V_{T} (e(k))}{T} }\, \, \le \\
&\ \ \ \ -\frac{\|e_{k}\|^{2}}{2\sqrt{1+\frac{\lambda _{\max }
(Q)}{\lambda _{\max } (P)}}+1} +T.\gamma _{c}^{2} \lambda _{\max }
(P)\left\| e_{k} \right\| ^{2} \label{Lyap3}.
\end{split}
\end{equation}
Now, we define the following functions:
\begin{align}
&{\alpha _{1} (\left\| e_{k} \right\| ) \triangleq \lambda _{\min }
(P)\left\| e_{k} \right\| ^{2} }\\
&{\alpha _{2} (\left\| e_{k}\right\| )\triangleq \lambda _{\max }
(P)\left\| e_{k} \right\| ^{2} } \\
&{\alpha_{3}(\left\|e_{k}\right\|)\triangleq\frac{1}{2\sqrt{1+\frac{\lambda
_{\max } (Q)}{\lambda _{\max }(P)}
} +1} \left\| e_{k} \right\| ^{2} } \\
&{\rho _{0} (T)\triangleq T}, \ {\gamma _{0} (\left\| e_{k} \right\|
)\triangleq \gamma _{c}^{2}
\lambda _{\max } (P)\left\| e_{k} \right\| ^{2} } \\
&{\gamma _{1}(\left\| x_{k} \right\| )\triangleq 0}, \ {\gamma
_{2}(\left\| u_{k} \right\| )\triangleq 0}.
\end{align}
Then, the following can be written
\begin{align}
&{\alpha _{1} (\left\| e_{k} \right\| )\le V_{T} (e(k))\le \,\alpha
_{2} (\left\| e_{k} \right\| )} \\
&\frac{V_{T} (e(k+1))-V_{T} (e(k))}{T} \le -\alpha _{3}
(\left\|e_{k}
\right\|)\notag\\
&\ \ \ \ +\rho _{0} (T)\left[\gamma _{0} (\left\| e_{k}
\right\|)+\gamma _{1} (\left\| x_{k} \right\| )+\gamma _{2} (\left\|
u_{k} \right\| )\right]
\end{align}
where $\alpha _{1} (.)$ , $\alpha _{2} (.)$ ,$\alpha _{3} (.)$ and
$\rho _{0} (.)$ are in class-$\mathcal{K}_{\infty}$ and $\gamma _{0}
(.)$ , $\gamma _{1} (.)$ and $\gamma _{2}(.)$ are nondecreasing
functions. Finally, since the trajectories of (\ref{dis3}) are
bounded for the Euler approximation, the Euler approximate model is
consistent with the exact model (\ref{dis1}). It follows that all
conditions of Theorem 1 in \cite{Arcak} are satisfied and the
proposed observer is semiglobal
practical in T. $\triangle$\\

\emph{\textbf{ Remark 3.}} $Q$ is not necessarily equal to {\it TI},
nevertheless, it can be figured out from (\ref{Lyap3}) that to have
a better convergence rate, $\lambda _{\max } (Q)$  must to be as
small as possible
\begin{equation}
\displaystyle \lambda _{\max } (Q)=\lambda _{\min }
(Q)=T\Leftrightarrow \, \, Q=TI.
\end{equation}

\emph{\textbf{ Remark 4: Nonlinear Uncertainty.}} Similar to Remark
2, in section 2, the observer is robust against any additive
Lipschitz nonlinear uncertainty with Lipschitz constant less than or
equal to $\gamma _{c}^{*} -\gamma _{c} $.


\section{Nonlinear $H_{\infty}$  Observer Synthesis}

In this section we extend the result of the previous section by
proposing a new nonlinear robust $H_{\infty}$ observer design
method. Consider the system
\begin{align}
x(k+1)&=A_{d}x(k)+F(x(k),u(k))+B_{d}w(k)\label{sys2_1}\\
y(k)&=C_{d}x(k)\label{sys2_2}
\end{align}
where $w(t)\in \ell_{2}[0,\infty)$ is an unknown exogenous
disturbance. Suppose that
\begin{equation}
z(k)=He(k)
\end{equation}
stands for the controlled output for error state where $H$ is a
known matrix. Our purpose is to design the observer parameter $L$
such that the observer error dynamics is asymptotically stable and
the following specified $H_{\infty}$ norm upper bound is
simultaneously guaranteed.
\begin{equation}
\|z\|\leq\mu\|w\|.
\end{equation}
The following theorem introduces a new method for nonlinear robust
$H_{\infty}$ observer design.\\

\emph{\textbf{Theorem 4.} Consider Lipschitz nonlinear system
(\ref{sys2_1})-(\ref{sys2_2}) with given Lipschitz constant
$\gamma_{d}$, along with the observer (\ref{observer1}). The
observer error dynamics is (globally) asymptotically stable with
minimum $\mathfrak{L}_{2}$ gain, $\mu^{\ast}$, if there exist
scalars $\epsilon> 0$ and $\zeta>0$, fixed matrix $Q>0$ and matrices
$P>0$ and $G$ such that the following LMI optimization problem has a
solution.}
\begin{equation}
\min(\zeta)\notag
\end{equation}
\begin{align}
&\left[
\begin{array}{cc}
P-Q-\varepsilon I & A_{d}^{ T} P-C_{d}^{T}G^{T} \\
PA_{d} -GC_{d} & P%
\end{array}
\right] >0 \label{H2}\\
&\left[
\begin{array}{cc}
\Psi_{1}I & P \\
P &  \Psi_{1}I %
\end{array}
\right] >0\label{H3} \\
&\left[
    \begin{array}{cc}
        \Lambda_{2}I  &  \frac{1}{2}\left[2(\gamma_{d}+1)\Psi_{1}+\lambda_{\max}(Q)\right]I \\
        \\\star  & B_{d}^{T}PB_{d}-\zeta I \\
    \end{array}%
\right]<0\label{H4}
\end{align}
where $\Psi_{1}$ is as in \eqref{si1} and
$\Lambda_{2}={H^{T}H}-Q+\gamma_{d}\left[3\Psi_{1}+\lambda_{\max}(Q)\right]$.\emph{Once
the problem is solved $L=P^{-1}G$ and
$\mu^{*}\triangleq\min(\mu)=\sqrt{\zeta}$.\\}

\textbf{Proof:} Consider the same Lyapunov function candidate as
before, thus,
\begin{equation}
\begin{split}
\Delta V =&\ e_{k}^{T} (A_{d} -LC_{d} )^{T} P(A_{d} -LC_{d} )e_{k} \\
&+2e_{k}^{T} (A_{d} -LC_{d} )^{T} P(F_{k} -\hat{F}_{k} ) -e_{k}^{T}Pe_{k}\\
&+(F_{k}-\hat{F}_{k})^{T}P(F_{k}-\hat{F}_{k})+2w_{k}^{T}B_{d}^{T}P(A_{d}-LC_{d})e_{k}\\
&+2w_{k}^{T}B_{d}^{T}P(F_{k}
-\hat{F}_{k})e_{k}+w_{k}^{T}B_{d}^{T}PB_{d}w_{k}\notag
\end{split}
\end{equation}
where the first four terms are the same as those found in Theorem
1, and the next three terms are due to the disturbance $w$. If
$w_{k}=0$, $\Delta V$ is given by (\ref{Lyap1}) so the LMIs
(\ref{H2}) and (\ref{H3}) guarantee the asymptotic stability. If
$w\neq 0$, we have that
\begin{align}
w_{k}^{T}B_{d}^{T}P(A_{d} -LC_{d})e_{k}&\leq
w_{k}^{T}B_{d}^{T}\bar{\sigma}(P)\bar{\sigma}(A_{d}
-LC_{d})e_{k}\notag\\
w_{k}^{T}B_{d}^{T}P(F_{k} -\hat{F}_{k})&\leq
w_{k}^{T}B_{d}^{T}\bar{\sigma}(P)\gamma_{d}e_{k}\notag
\end{align}
from the above and using (\ref{f1}), (\ref{f2}) and (\ref{ineq6}),
we have:
\begin{equation}
\begin{split}
\Delta V \leq &e_{k}^{T}\left[-Q+ \gamma_{d}\left(3\lambda_{\max}(P)+\lambda_{\max}(Q)\right)\right] e_{k} \\
&+w_{k}^{T}B_{d}^{T}\left[2\lambda_{\max}(P)(\gamma_{d}+1)+\lambda_{\max}(Q)\right]e_{k}\\
&+w_{k}^{T}B_{d}^{T}PB_{d}w_{k}.
\end{split}
\end{equation}
Now, define
\begin{equation}
J\triangleq
\sum_{k=0}^{\infty}\left[z(k)^{T}z(k)-\mu^{2}w(k)^{T}w(k)\right].
\end{equation}
So, $J <
\sum_{k=0}^{\infty}\left[z(k)^{T}z(k)-\mu^{2}w(k)^{T}w(k)+\Delta
V\right]$. Thus, a sufficient condition for $J\leq0$ is that
\begin{equation}
\forall \hspace{1mm} k \in[0,\infty),\hspace{5mm}
z^{T}z-\mu^{2}w^{T}w+\Delta V\leq0.
\end{equation}
We have
\begin{gather}
\begin{split}
z(k)^{T}z(k)&-\mu^{2}w(k)^{T}w(k)+\Delta V \leq e_{k}^{T}[H^{T}H-Q\cdots\\
&\cdots+\gamma_{d}\left(3\lambda_{\max}(P)+\lambda_{\max}(Q)\right)]e_{k}\\
&+w_{k}^{T}B_{d}^{T}\left[2\lambda_{\max}(P)(\gamma_{d}+1)+\lambda_{\max}(Q)\right]e_{k}\\
&+w_{k}^{T}(B_{d}^{T}PB_{d}-\mu^{2}I)w_{k}. \label{H1}
\end{split}
\end{gather}
So a sufficient condition for $J\leq0$ is that the right hand side
of the above inequality be negative. Then
\begin{equation}
z^{T}z-\mu^2\ w^{T}w\leq0\Rightarrow\|z\|\leq\mu|w\|
\end{equation}
substituting $\lambda_{\max}(P)$ from (\ref{lambda2}) into
(\ref{H1}) and defining $\zeta=\mu^2$, the LMI (\ref{H4}) is
obtained. $\triangle$\\

\emph{\textbf{Remark 5.}} For the Euler approximation, according to
Theorem 3, if $\lambda_{\min}(Q)=T$ then the proposed $H_{\infty}$
observer will be semiglobal practical in $T$. In this case, if we
choose $Q=TI$ as suggested in Remark 3, then it is clear that LMIs
(\ref{H3}) and (\ref{H4}) can be simplified. Furthermore, having
$\gamma_{d}<1$, we can first maximize the admissible Lipschitz
constant using Theorem 2, and then minimize $\mu$ for the maximized
$\gamma_{d}$, using Theorem 4. In this case, according to Remark 2,
robustness against nonlinear uncertainty is also guaranteed.\\

Now we show the usefulness of this method through a design
example.\\

\emph{\textbf{Example:}} Consider the following continuous-time
nonlinear system and its Euler approximation
\begin{equation}
\begin{array}{l} {\dot{x}=\left[\begin{array}{cc} {0} &
{1} \\ {-1} & {-1} \end{array}\right]x+\left[\begin{array}{c}
{x_{1}^{3} }
\\ {-6x_{1}^{5} -6x_{1}^{2} x_{2} -2x_{1}^{4} -2x_{1}^{3} }
\end{array}\right]} \\
{y=Cx=\left[\begin{array}{cc} {1} & {0}
\end{array}\right]x} \\
{x^{a} (k+1)=(I+AT)x^{a} (k)+Tf(x^{a} (k))}
\\ {y(k)=\left[\begin{array}{cc} {1} & {0}
\end{array}\right]x^{a} (k)}\notag.
\end{array}
\end{equation}
It is well-known that the polynomial type nonlinearities are locally
Lipschitz. $f(x)$ is Lipschitz in the following region
\begin{equation}
\mathcal{D}=\biggl\lbrace (x_{1},x_{2})\in \mathbb{R}^{2} \ | \
x_{1}\leq 0.3 \biggr\rbrace \notag
\end{equation}
with Lipschitz constant $\gamma_{c}=0.6109$. We assume  $T=0.1$ sec
and design observer (\ref{observer2}). Using Theorem 2, we have
\begin{align}
\gamma _{c}^{*}=0.67\notag.
\end{align}
Now, using Theorem 4, with $H=0.25I, \
\gamma_{d}=\gamma_{d}^{*}=T\gamma_{c}^{*}, \ B=\left[
       \begin{array}{cc}
       1 & 1 \\
       \end{array}
\right]^{T}$, we get
\begin{align}
\mu^{*}=0.1308, \ L=\left[
\begin{array}{cc}
  1.0497 & 0.3588
\end{array}
\right]^{T}\notag.
\end{align}
Figure 1, shows the state trajectories for the continuous-time
system along with their estimations made by an observer which uses
the Euler-approximate model. Simulation is done for 10 seconds (100
samples) in the presence of a Gaussian disturbance with zero mean and standard deviation $0.01$. 
It can be seen in the figure that after 3 seconds, the true and
estimated states are almost identical.

\begin{figure}[!h]
  \centering
  \includegraphics[width=3.5in]{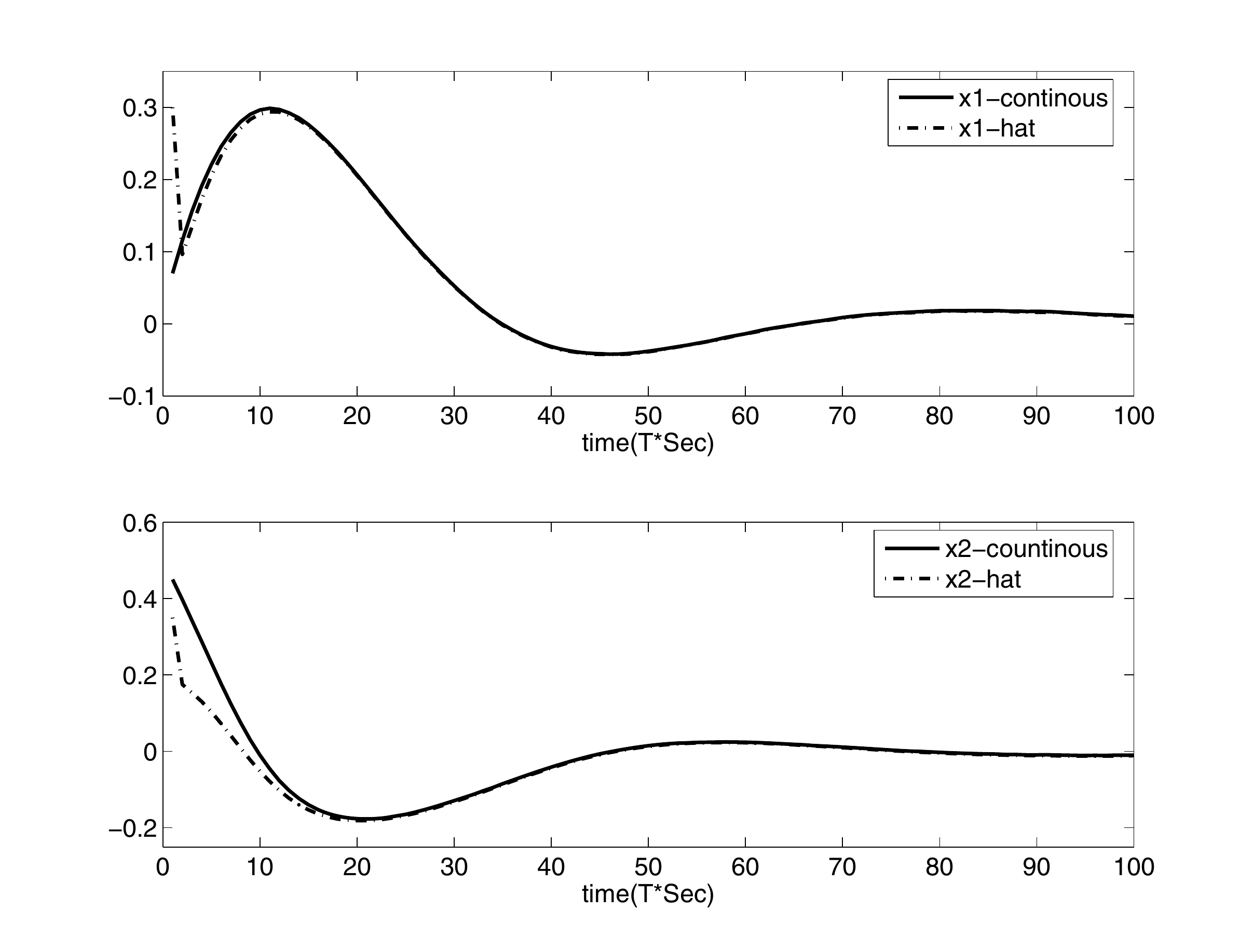}\\
  \caption{The true and estimated states} \label{Fig1}
\end{figure}

\subsection{Higher Order Approximate Models}
Under certain conditions, the Lipschitz contiguity is preserved under second order approximate discretization as studied in \cite{abbaszadeh2016lipschitz}.
In most practical applications, first or second order discretization should be enough, specially since the sampling time can be selected small enough to ensure desired bounds on the approximation error. Furthermore, the expressions involving higher-order approximate models rapidly become very complicated. In particular, higher-order partial derivatives require tensor analysis of higher-orders.
Under the ZOH assumption, similar to the approach given in \cite{Kazantzis1999763}, we have:
\begin{equation}
\begin{split}
x(k+1)&=x(k)+{\sum_{l=1}^{\infty}\frac{T^{l}}{l!}\frac{d^{l}x}{dt^{l}}}|_{t_{k}}\\
&=x(k)+\sum_{l=1}^{\infty}\frac{T^{l}}{l!}\frac{d^{l-1}}{dt^{l-1}}[Ax+f(x,u)]|_{t_{k}}\\
&=x(k)+\sum_{l=1}^{\infty}\frac{T^{l}}{l!}[A\frac{d^{l-1}x}{dt^{l-1}}+\frac{d^{l-1}}{dt^{l-1}}f(x,u)]|_{t_{k}}.
\end{split}
\end{equation}
where {\large\begin{equation} \left\{
  \begin{array}{l}
    \frac{d}{dt}f(x,u)=\frac{\partial f}{\partial x}\cdot
\frac{dx}{dt}+\frac{\partial f}{\partial u}\cdot \frac{du}{dt}\\
    \frac{d^{n}}{dt^{n}}f(x,u)=
\frac{d}{dt}[\frac{d^{n-1}}{dt^{n-1}}f(x,u)], \ \ \ n \geq 2 \\
  \end{array}
\right.
\end{equation}}
Under the ZOH assumption, $\frac{du}{dt}=0$ in each sampling
interval and thus:
\begin{equation}
\begin{split}
x(k+1)&=x(k)+\sum_{l=1}^{\infty}\frac{T^{l}}{l!}[A\frac{d^{l-1}x}{dt^{l-1}}+\frac{d^{l-1}}{d
t^{l-1}}f(x,u)]|_{t_{k}} \\
\frac{d}{dt}f(x,u)&=\frac{\partial f}{\partial x}\cdot
\frac{dx}{dt}, \ \ \ \ \ \ \ \frac{d^{n}f}{dt^{n}}=
\frac{d}{dt}(\frac{d^{n-1}f}{dt^{n-1}}), \ n \geq 2. \label{taylor2}
\end{split}
\end{equation}
The first order approximation, ($l = 1$) leads to the well-known Euler
approximate model.

The robust nonlinear observer design results can be extended to higher-order approximate models as well, which is a topic for further research.


\section{Conclusion}
In this paper, a new algorithm for robust $H_{\infty}$ nonlinear
observer design for nonlinear discrete-time systems was proposed
based on an LMI approach. The observer is robust in the sense that
it can achieve convergence to the true state, despite  nonlinear
model uncertainty with guaranteed exogenous disturbance rejection
ratio. In addition, when the exact discrete-time model of the
system is not available, the same algorithm can still be used for
the Euler approximated model. In the proposed algorithms, the
admissible Lipschitz constant and the disturbance attenuation
level can be maximized through LMI optimization. These features
make the proposed algorithm an efficient design method.

\bibliographystyle{IEEEtran}
\bibliography{References_auto}

\end{document}